\documentclass[10pt,conference]{IEEEtran}
\IEEEoverridecommandlockouts
\usepackage{cite}
\usepackage{amsmath,amssymb,amsfonts}
\usepackage{algorithmic}
\usepackage{graphicx}
\usepackage{textcomp}
\usepackage{xcolor}
\usepackage{citesort}
\usepackage{subfigure}
\usepackage[justification=centering]{caption}

\newtheorem{theorem}{Theorem}

\newtheorem{lemma}{Lemma}

\newtheorem{corollary}{Corollary}

\newtheorem{proposition}{Proposition}

\usepackage{algorithm}
\usepackage{algorithmic}
\usepackage{cases}
\usepackage[
top    = 1.80cm,
bottom = 1.05in,
left   = 0.63 in,
right  = 0.63 in]{geometry}

\begin{document}
\title{Reconfigurable Intelligent Surface Assisted NOMA Empowered Integrated Sensing and Communication\\
}
\author{
	\IEEEauthorblockN{
		Jiakuo~Zuo\IEEEauthorrefmark{1} and 
		Yuanwei~Liu\IEEEauthorrefmark{2} 
		%Zhiguo~Ding\IEEEauthorrefmark{3}, and Lingyang~Song\IEEEauthorrefmark{4} 
	}
	\IEEEauthorblockA{
		\IEEEauthorrefmark{1} School of Internet of Things, Nanjing University of Posts and Telecommunications, Nanjing 210003, China.\\
		%\IEEEauthorrefmark{1} Jiangsu Key Laboratory of Broadband Wireless Communication and Internet of Things,\\ Nanjing University of Posts and Telecommunications, Nanjing 210003, China.\\
		%\IEEEauthorrefmark{1} Key Laboratory of Universal Wireless  Communications (BUPT), Ministry of Education, Beijing 100876, China.\\
	 	\IEEEauthorrefmark{2} School of Electronic Engineering and Computer Science, Queen Mary University of London, London E1 4NS, U.K.\\
		%	\IEEEauthorrefmark{3} School of Electrical and Electronic Engineering, University of Manchester, Manchester, U.K.\\
		%	\IEEEauthorrefmark{4} Department of Electronics, Peking University, Beijing 100871 China. \\
		E-mail: zuojiakuo@njupt.edu.cn, yuanwei.liu@qmul.ac.uk. 
		  \vspace{-1cm}
		  %zhiguo.ding@manchester.ac.uk, lingyang.song@pku.edu.cn.
	} 
	\thanks{This work was supported by Key Laboratory of Universal Wireless Communications (BUPT), Ministry of Education, P.R.China under Grant KFKT-2022105 and China Postdoctoral Science Foundation under Grant 2021M693699.}
}
\maketitle
 \begin{abstract}
     This paper exploits the potential of reconfigurable intelligent surface (RIS) to improve radar sensing in a non-orthogonal multiple access (NOMA) empowered integrated sensing and communication (NOMA-ISAC) network. The objective is to maximize the minimum radar beampattern gain by jointly optimizing the active beamforming, power allocation coefficients and passive beamforming. To tackle the formulated non-convex problem, we propose an efficient joint optimization algorithm by invoking alternating optimization, successive convex approximation (SCA) and sequential rank-one constraint relaxation (SRCR) algorithm. Numerical results show that the proposed RIS assisted NOMA-ISAC system, with the aid of the proposed scheme, outperforms the RIS assisted ISAC system without NOMA. %It is also demonstrated that the proposed joint optimization scheme can enhances the system performance compared to varous benchmark schemes.
     %scheme over various benchmark schemes.
 \end{abstract}
  
   \vspace{-0.5cm}
\section{Introduction}
In conventional wireless communication, communication quality of service (QoS) is the main indicator. However, in the upcoming beyond fifth generation (B5G) and sixth generation (6G) wireless networks, sensing service will play a more important role than ever before~\cite{9585321}. The goal of future wireless communication is to provide services based on both sensing and communication functionalities. Toward this trend, integrated sensing and communication (ISAC) has emerged and attracted growing attention in both academia and industries. %Due to the numerous advantages offered by ISAC, there has been a number of research contributions recently [xxxx]. 
 %Sensing as A Service in 6G Perceptive Networks: A  Unified Framework for ISAC Resource Allocation
%The ISAC can be performed with the aid of various key enabling technologies

To boost the communication and sensing capabilities, various key technologies can be exploited in ISAC, such as: millimeter wave~\cite{Gao2022IntegratedSA}, terahertz~\cite{Chaccour2021JointSA} and massive multiple-input
multiple-output (MIMO)~\cite{You2022BeamSI}. However, the communication users will suffer from severe interference when the system is overloaded, which is not considered in these works. Non-orthogonal multiple access (NOMA) is a good solution as it can multiplex communication users and mitigate the inter-user interference.
%by improving the spectral efficiency and spatial degrees of freedom (DoF),
%The ISAC can be performed with the aid of various key enabling technologies, including the millimeter wave (mmWave), ultra-dense network (UDN) and multiple-input-multiple-output (MIMO) radar. These technologies have been incorporated to boost the communication and sensing capabilities by improving the spectral efficiency and spatial degrees of freedom (DoF), but are still unable to adequately address several critical challenges
The authors in~\cite{Wang2022NOMAEI} first proposed a NOMA empowered ISAC (NOMA-ISAC) system, where the superimposed NOMA communication signal was utilized to perform radar sensing. 
% NOMA Empowered Integrated Sensing and Communication
Considering that the sensing signal can also be information-bearing, the authors in~\cite{Wang2021NOMAII} used multiple beams of the sensing signal to deliver extra multicast information and detect radar targets simultaneously. %The communications users received one desired unicast stream and multiple multicast streams, which were detected by the successive interference cancellation (SIC).
% NOMA Inspired Integrated Sensing and Communication
% NOMA Inspired Interference Cancellation for Integrated Sensing and Communication
%In conventional ISAC networks, the total bandwidth was assumed to be used for both radar detection and wireless communication. However, this assumption is unpractical. Because the bandwidth has been occupide by different applications. 
The authors in~\cite{Mu2022NOMAAidedJR} proposed a novel NOMA aided joint radar and multicast-unicast communication system, where the base station (BS) transmits superimposed multicast and unicast messages to the radar users and communication users, while detecting the radar user targets. %The C-users acted as the NOMA "strong" users and the R-users acted as the NOMA "weak" users.
% NOMA-Aided Joint Radar and Multicast-Unicast Communication Systems
Since the bandwidth has been occupied by different applications, the ISAC networks can not utilize all the spectrum for both radar detection and wireless communication. To overcome this difficulty, the authors in~\cite{Zhang2022SemiIntegratedSensingandCommunicatios} proposed an uplink NOMA-assisted semi-ISAC system, where the total bandwidth was divided into ISAC bandwidth and communication-only bandwidth.
%Semi-Integrated-Sensing-and-Communication (Semi-ISaC): From OMA to NOMA

In conventional ISAC systems, radar sensing relies on line-of-sight (LoS) links between the BS and radar targets. 
% assume that there are line-of-sight (LoS) links between the BS and radar targets. 
However, in practical scenarios, the radar targets are likely to be distributed in the non-LoS (NLoS) region of the BS. As a remedy, reconfigurable intelligent surface (RIS) can be integrated into ISAC systems.
%to improve the sensing and communication coverage and enhance the system performance. 
RIS comprises a large number of passive elements and can reconfigure the signal propagation by adjusting the phase shifts. Also, the RIS can provide virtual LoS links to the ISAC systems, potentially resulting in better sensing performance~\cite{Song2022JointTA}. %The authors in~\cite{Song2022JointTA} studies the joint active and passive beamforming problem for RIS asisted ISAC (RIS-ISAC) system, where the radar targets are at the NLoS areas of the BS.

%and each element can independently reflect the incident signal by adjusting the phase shifts, so that the signal propagation between the BS and uers/radar targets can be reconfigured. Also, the RIS 

Inspired by the aforementioned discussion, we investigate the RIS assisted NOMA-ISAC (RIS-NOMA-ISAC) system. To the best of our
knowledge, the joint optimization design for the considered system has not been studied yet. In this paper, the communication users and radar targets are assumed to be at the NLoS areas of the BS. The RIS is deployed to create reflection and virtual LoS links for communication and target sensing. A novel alternating algorithm is proposed to optimize the active beamforming, power allocation coefficients and passive beamforming.

%wireless sensing relies on line-of-sight (LoS) links between the access points (APs) (or sensing transceivers) and the sensing targets, such that the sensing information (e.g., the distance/velocity/angle of targets) can be extracted based on the target echo signals. However, in practical scenarios with dense obstructions, the sensing targets are highly likely to be located at the non-LoS (NLoS) region of APs, where conventional LoS sensing is not applicable in general. Therefore, how to realize NLoS sensing in such scenarios is a challenging task.
% [Itelligent Reflecting Surface Enabled Sensing: Cram´er-Rao Lower Bound Optimization]
 %The rest of this paper is organized as follows. In Section II, the system model and the problem formulation for designing the RIS-CNOMA system are presented. In Sections III and IV, we propose the alternating optimization based optimal algorithm and the low-complexity alternating optimization based suboptimal algorithm to solve the original optimization problem, respectively. Numerical results are presented in Section V, which is followed by the conclusions in Section VI.

Notations: $\mathcal{C} ^{M\times 1}$ denotes a complex vector of size \emph{M}, diag(\textbf{x}) denotes a diagonal matrix whose diagonal elements are the corresponding elements in vector \textbf{x}. The $(m,n)$-th element of matrix $\textbf{X}$ is denoted as $\left[ \mathbf{X} \right] _{m,n}$. ${\textbf{x}}^{H}$ and ${\textbf{X}}^{H}$ denote the conjugate transpose of vector \textbf{x} and matrix \textbf{X}, respectively. The notations Tr(\textbf{X}) and rank(\textbf{X}) denote the trace and rank of matrix \textbf{X}, respectively. $\mathcal{C}\mathcal{N}\left( 0,\sigma ^2 \right) $ represents a random complex variable following the distribution of zero mean and $\sigma ^2$ variance.
 \vspace{-0.3cm}

 \section{System Model and Problem Formulation}
  \begin{figure}[t]
 	 \setlength{\abovecaptionskip}{3pt}
 	 \setlength{\belowcaptionskip}{-20pt}
 	\centering
 	\includegraphics[scale=0.15]{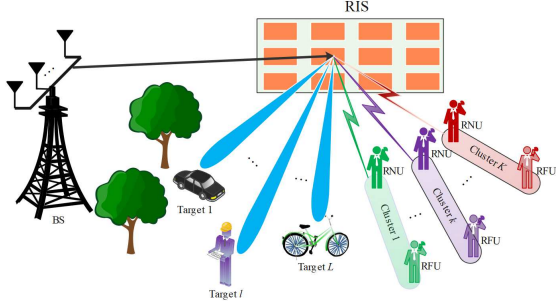}
 	\caption{System model of the RIS-NOMA-ISAC system}
 	\label{ST1}
 \end{figure}
As shown in Fig.~\ref{ST1}, the considered RIS-NOMA-ISAC system consists of a dual-functional base station (BS) equipped with $N_T$  antennas, $2K$ single-antenna users, an uniform linear array (ULA)-RIS with $M$ reflecting elements and $L$ radar targets. We assume that the direct links from the BS to the communication users and radar targets are blocked, which is a typical application scenario when RISs are needed or used. Therefore, in our system, we assume that all the communication users and radar targets are at the NLoS areas of the BS. We carefully deploy the RIS to ensure that all the targets are in LoS areas of the RIS. 

To improve the spectral efficiency and reduce the system load, we assume that the $2K$ users are grouped into $\emph{K}$ clusters by employing user clustering techniques. As a result, there are two types of users 
in each cluster, namely the RIS-near user (RNU) and the RIS-far user (RFU). Generally, the
RNUs are much closer to the RIS than the RFUs.  In each
cluster, the NOMA protocol is applied during transmission.
The cluster and user sets are denoted by $\mathcal{K} =\left\{ 1,\cdots ,K \right\}$ and $\mathfrak{U} =\left\{ 1,\cdots ,2K \right\}$, respectively. Moreover, denote by ${{\mathfrak U}_k}$ the set of users in cluster $k$, where $\mathfrak{U} =\cup _{k\in} \mathfrak{U}  _k$, $ \mathfrak{U}  _k\cap  \mathfrak{U}  _{\underline{k}}=\oslash \left( k,\underline{k}\in  \mathcal{K} ,k\ne \underline{k} \right) $. For notation simplicity, we denote the RNU and RFU in the $k$-th cluster as user $\mathcal{U}\left( k,n \right) $ and user $\mathcal{U}\left( k,f \right) $, respectively. 
\vspace{-0.2cm}
\subsection{Communication Model}
\vspace{-0.2cm}
The superimposed communication signal transmitted by the BS is given by  
 \begin{equation}\label{superimposed_signal}
 	 \setlength{\abovedisplayskip}{0pt}
 	 \setlength{\belowdisplayskip}{0pt}
	\mathbf{x}=\sum_{k=1}^K{\mathbf{w}_k\left( \sqrt{a _{k,n}}s_{k,n}+\sqrt{a _{k,f}}s_{k,f} \right)}, 
 \end{equation}
 where $\mathbf{w}_k\in \mathcal{C}^{N_{\text{T}}\times 1}$ is the active beamforming vector for the $k$-th cluster, $s_{k,i}$ denotes the communication signal to be sent to user $\mathcal{U}\left( k,i \right) $ with $\mathbb{E}\left( s_{k,i}^{H}s_{k,i} \right) =1$, $a _{k,i}$ is the corresponding power allocation coefficient, $i\in \left\{ n,f \right\} $, $k\in \mathcal{K}$.
%We further assume that all the perfect CSI are available at the BS. $\footnote{The channel acquisition methods for the RIS aided CNOM empowered ISAC system are outside the scope of this work. However, our results can serve as a theoretical system performance benchmark. In addition, channel estimation methods~\cite{9130088,9366805} proposed for conventional RISs can be applied for STAR-RISs. The design of more efficient CSI estimation methods and analysis of imperfect CSI for RIS aided CNOM empowered ISAC systems are interesting but challenging new research problems.}$
Let $\mathbf{G}\in \mathcal{C}^{M\times N_{\text{T}}}$ and $\mathbf{g}_{k,i}\in \mathcal{C}^{M\times 1}$ be the channel coefficients of the communication links $\text{BS}\rightarrow \text{RIS}$ and $\text{RIS}\rightarrow \mathcal{U}\left( k,i \right)$, respectively. With the help of RIS, the signal received at user $\mathcal{U}\left( k,i \right) $ can be mathematically expressed as
 \begin{equation}\label{signal_yki}
  	\setlength{\abovedisplayskip}{0pt}
  	\setlength{\belowdisplayskip}{0pt}
 	y_{k,i}=\left( \mathbf{g}_{k,i}^{H}\mathbf{\Theta G} \right) \sum_{k=1}^K{\mathbf{w}_k\left( \sqrt{a_{k,n}}s_{k,n}+\sqrt{a_{k,f}}s_{k,f} \right)} + z_{k,i},
\end{equation}
where $\mathbf{\Theta }=\mathrm{diag}\left( \mathbf{v} \right) $ is the RIS's diagonal phase shift matrix, $\mathbf{v}=\left[ e^{j\theta _{1}^{\text{RIS}}}e^{j\theta _{2}^{\text{RIS}}}\cdots e^{j\theta _{M}^{\text{RIS}}} \right] 
$ is the passive beamforming vector of the RIS with $\theta _{m}^{\text{RIS}}\in \left[ 0,2\pi \right) $ denoting the phase shift of the $m$-th reflecting element, $m\in \mathcal{M} 
$, and $z_{k,i}\sim \mathcal{C} \mathcal{N} \left( 0,\sigma ^2 \right) $ is the additive white Gaussian noise (AWGN), $\mathcal{M}=\left\{ 1,2,\cdots ,M \right\} $ is the reflecting elements set.

In the proposed RIS-NOMA-ISAC system, successive interference cancellation (SIC) is applied for the users to decode their signals in the same cluster. We assume a fixed decoding order in each cluster. Particularly, user $\mathcal{U}\left( k,n \right) $ first decodes the signal of user $\mathcal{U}\left( k,f \right) $ and then subtracts this signal from its observation to decode its own information. Therefore, the achievable rate for user $\mathcal{U}\left( k,n \right) $ to decode the signal of user $\mathcal{U}\left( k,f \right) $ is given by
 \begin{equation}\label{Rn21} 
   	\setlength{\abovedisplayskip}{-1pt}
  	\setlength{\belowdisplayskip}{-3pt}
 	R_{k,f\rightarrow n}=\log _2\left( 1+\frac{a_{k,f}\left| \mathbf{g}_{k,n}^{H}\mathbf{\Theta Gw}_k \right|^2}{I_{k,n}^{\mathrm{iner}}+I_{k,n}^{\mathrm{iter}}+\sigma ^2} \right),  
\end{equation} 
where $I_{k,n}^{\mathrm{iner}}=a_{k,n}\left| \mathbf{g}_{k,n}^{H}\mathbf{\Theta Gw}_k \right|^2
$ and $I_{k,n}^{\mathrm{iter}}=\sum_{\widetilde{k}\ne k}^K{\left| \mathbf{g}_{k,n}^{H}\mathbf{\Theta Gw}_{\widetilde{k}} \right|^2}
$.

 If the above decoding is successful, then user $\mathcal{U}\left( k,n \right) $ removes the signal $s_{k,f}$ from $y_{k,n}$ to further decode its own signal $s_{k,n}$. The corresponding individual achievable rate is given by
 \begin{equation}\label{Rn1} 
 	   	\setlength{\abovedisplayskip}{-1pt}
 	\setlength{\belowdisplayskip}{-3pt}
	R_{k,n}=\log _2\left( 1+\frac{a_{k,n}\left| \mathbf{g}_{k,n}^{H}\mathbf{\Theta Gw}_k \right|^2}{I_{k,n}^{\mathrm{iter}}+\sigma ^2} \right).    
\end{equation} 

Accordingly, user $\mathcal{U}\left( k,f \right) $ directly decodes its own signal by treating the signal of user $\mathcal{U}\left( k,n \right) $ as interference. Therefore, the achievable rate for user $\mathcal{U}\left( k,f \right) $ to decode its own signal can be expressed as
 \begin{equation}\label{Rn22}
 	   	\setlength{\abovedisplayskip}{-3pt}
 	\setlength{\belowdisplayskip}{-3pt}
 	R_{k,f\rightarrow f}=\log _2\left( 1+\frac{a_{k,f}\left| \mathbf{g}_{k,f}^{H}\mathbf{\Theta Gw}_k \right|^2}{I_{k,f}^{\mathrm{iner}}+I_{k,f}^{\mathrm{iter}}+\sigma ^2} \right),  
\end{equation} 
where $I_{k,f}^{\mathrm{iner}}=a_{k,n}\left| \mathbf{g}_{k,f}^{H}\mathbf{\Theta Gw}_k \right|^2
$ and $I_{k,f}^{\mathrm{iter}}=\sum_{\widetilde{k}\ne k}^K{\left| \mathbf{g}_{k,f}^{H}\mathbf{\Theta Gw}_{\widetilde{k}} \right|^2}$.
 
 As a result, the individual achievable rate of user $\mathcal{U}\left( k,f \right) $ is given by
 \begin{equation}\label{Rn2}
   \setlength{\abovedisplayskip}{-5pt}
  \setlength{\belowdisplayskip}{-5pt}
	R_{k,f}=\min \left\{ R_{k,f\rightarrow n},R_{k,f\rightarrow f} \right\}.   
\end{equation}  
\vspace{-1cm} 
\subsection{Radar Detection Model}
%\vspace{-0.5cm} 
 Since all the potential targets are in the NLoS ares of the BS, NLoS links from the BS can be exploited to perform radar target sensing. However, we can utilize the reflection-LoS links created by the RIS to complete the sensing tasks. We consider using the RIS's beampattern gain as the sensing performance metric. The reflected signal at the RIS can be expressed as
  \begin{equation}\label{signal_RIS} 
  	   \setlength{\abovedisplayskip}{-3pt}
  	\setlength{\belowdisplayskip}{-3pt}
 	\overline{\mathbf{x}}=\mathbf{\Theta G}\left( \sum_{k=1}^K{\mathbf{w}_k\left( \sqrt{a _{k,n}}s_{k,n}+\sqrt{a _{k,f}}s_{k,f} \right)}  \right). 
 \end{equation}
 
 Therefore, the corresponding covariance matrix is given by
 \begin{equation}\label{covariance matrix}  
 	   \setlength{\abovedisplayskip}{0pt}
 	\setlength{\belowdisplayskip}{0pt}
	\mathbf{R}_{\overline{\mathbf{x}}}=\mathbb{E}\left( \overline{\mathbf{x}}\overline{\mathbf{x}}^H \right) =\mathbf{\Theta G}\left( \sum_{k=1}^K{\mathbf{w}_k\mathbf{w}_{k}^{H}}  \right) \mathbf{G}^H\mathbf{\Theta }^H.
 \end{equation}

 In our considered system, the communication signal is used to perform radar target sensing, thus the RIS's beampattern gain with respect
 to the $q$-th interested angle, i.e,  $\theta _{q}^{\text{Tg}}$, is given by
 \begin{equation}\label{beam pattern}  
  \setlength{\abovedisplayskip}{0pt} 
	\mathcal{P}_{\theta _{q}^{\text{Tg}}}\left( \mathbf{w}_k,\mathbf{v} \right) =\boldsymbol{a }^H\left( \theta _{q}^{\text{Tg}} \right) \mathbf{\Theta G}\left( \sum_{k=1}^K{\mathbf{w}_k\mathbf{w}_{k}^{H}}  \right) \mathbf{G}^H\mathbf{\Theta }^H\boldsymbol{a }\left( \theta _{q}^{\text{Tg}} \right), 
 \end{equation} 
where $\boldsymbol{a }\left( \theta \right) =\left[ 1,e^{j\frac{2\pi d}{\lambda}\sin \left( \theta \right)},\cdots ,e^{j\frac{2\pi d}{\lambda}\left( M-1 \right) \sin \left( \theta \right)} \right] ^T$ is the steering vector at the RIS with angle $\theta$, $d$ denotes the antenna spacing, $\lambda $ denotes the carrier wavelength, $q\in \mathcal{Q} \triangleq \left\{ 1,2,\cdots ,Q \right\} 
$, and $\mathcal{Q} _{\theta}=\left\{ \theta _{1}^{\mathrm{Tg}},\theta _{2}^{\mathrm{Tg}},\cdots ,\theta _{Q}^{\mathrm{Tg}} \right\} 
$ is the set of interested sensing angles. 
\subsection{Maximize The Minimum Beampattern Gain} 
The objective of this paper is to maximize the minimum beampattern gain towards \textit{Q} interested angles by jointly
optimizing the active beamforming and power allocation coefficients at the BS, and the passive beamforming at the RIS.% subject to the minimum QoS required by communication users and the maximum total transmit power constraint. 
~Accordingly, the optimization problem is formulated as
\begin{subequations}\label{OP_MM}
	 \setlength{\abovedisplayskip}{-5pt}
	 \setlength{\belowdisplayskip}{0pt}
	\begin{align}
	&\underset{\mathbf{a}_k,\mathbf{w}_k,\mathbf{v}}{\max}\underset{q\in \mathcal{Q}}{\min}\mathcal{P}_{\theta _{q}^{\text{Tg}}}\left( \mathbf{w}_k,\mathbf{v} \right) , \\
	&s.t.~R_{k,i}\geqslant R_{k,i}^{\min}, \label{OP_MM:b} \\
	&   \ \ \ \ \   \sum_{k=1}^K{\lVert \mathbf{w}_k \rVert _2}   \leqslant P_{\max}
	, \label{OP_MM:c}  \\
	&   \ \ \ \ \  a _{k,n}+a _{k,f}=1, a_{k,i}\in \left( 0,1 \right) , \label{OP_MM:d}  \\	
	&   \ \ \ \ \   \theta _{m}^{\text{RIS}}\in \left[ 0,2\pi \right)  ,\label{OP_MM:e} 
\end{align} 
\end{subequations}   
 where $ R_{k,i}^{\min}$ is the minimum QoS requirement, $P_{\max}$ is the maximum transmit power at the BS, $\mathbf{a}_k=\left[ a_{k,n}a_{k,f} \right] ^T$, $i\in \left\{ n,f \right\}, k\in \mathcal{K}$, $m\in \mathcal{M}$.
 \section{Proposed Solution}
 In this section, we elaborate on how to solve problem~
 \eqref{OP_MM}. We first transform problem~
 \eqref{OP_MM} into a more tractable form. To facilitate the design, we define $\mathbf{W}_k=\mathbf{w}_k\mathbf{w}_{k}^{H}$, where $\mathbf{W}_k\succcurlyeq 0$ and $\text{rank}\left( \mathbf{W}_k \right) =1$, $k\in \mathcal{K}$. Similarly, we define $\mathbf{V}=\mathbf{vv}^H
 $,  which satisfies $\mathbf{V}\succcurlyeq 0$ and $\text{rank}\left( \mathbf{V} \right) =1
 $. Then, the beampattern gain in~\eqref{beam pattern} can be rewritten as:
 \begin{equation}\label{beam pattern rewritten}  
 		 	\setlength{\abovedisplayskip}{0pt}
 	\setlength{\belowdisplayskip}{0pt}
	\mathcal{P}_{\theta _{q}^{\text{Tg}}}\left( \mathbf{W}_k,\mathbf{V} \right) =\text{Tr}\left[ \mathbf{V\Upsilon }_q\left( \sum_{K=1}^K{\mathbf{W}_k}  \right) \mathbf{\Upsilon }_{q}^{H} \right] , 
 \end{equation} 
 where $\mathbf{\Upsilon }_q=\mathrm{diag}\left\{ \boldsymbol{a }^H\left( \theta _{q}^{\text{Tg}} \right) \right\} \mathbf{G}$.% $q\in \mathcal{Q}$.
 
 Moreover, the quadratic terms $\left| \mathbf{g}_{k,i}^{H}\mathbf{\Theta Gw}_k \right|^2
 $ and $\left| \mathbf{g}_{k,i}^{H}\mathbf{\Theta Gw}_{\widetilde{k}} \right|^2
 $ in~\eqref{Rn21}, ~\eqref{Rn1} and~\eqref{Rn22} can be rewritten as:
  \begin{equation}\label{channel gain rewritten}
    \begin{cases}
		\left| \mathbf{g}_{k,i}^{H}\mathbf{\Theta Gw}_k \right|^2=\text{Tr}\left( \mathbf{V\Gamma }_{k,i}\mathbf{W}_k\mathbf{\Gamma }_{k,i}^{H} \right),\\
		\left| \mathbf{g}_{k,i}^{H}\mathbf{\Theta Gw}_{\widetilde{k}} \right|^2=\text{Tr}\left( \mathbf{V\Gamma }_{k,i}\mathbf{W}_{\widetilde{k}}\mathbf{\Gamma }_{k,i}^{H} \right),\\
	\end{cases} 
 \end{equation} 
 where $\mathbf{\Gamma }_{k,i}=\mathrm{diag}\left\{ \mathbf{g}_{k,i}^{H} \right\} \mathbf{G}$, $k\ne \tilde{k}
 $.%, $i\in \left\{ n,f \right\}$, $k\in \mathcal{K}$.
 
By exploiting the above definitions, constraints in~\eqref{OP_MM:b} can be equivalently reformulated as follows:
 \begin{equation}\label{R11 constrait} 
 	a_{k,n}\mathrm{Tr}\left( \mathbf{V\Gamma }_{k,n}\mathbf{W}_k\mathbf{\Gamma }_{k,n}^{H} \right) \geqslant r_{k,n}^{\min}\left( I_{k,n}^{\mathrm{iter}}+\sigma ^2 \right), 
 \end{equation}  
 \begin{equation}\label{R21 constrait} 
	a_{k,f}\mathrm{Tr}\left( \mathbf{V\Gamma }_{k,n}\mathbf{W}_k\mathbf{\Gamma }_{k,n}^{H} \right) \geqslant r_{k,f}^{\min}\left( I_{k,n}^{\mathrm{iner}}+I_{k,n}^{\mathrm{iter}}+\sigma ^2 \right), 
  \end{equation} 
 \begin{equation}\label{R22 constrait} 
	a_{k,f}\mathrm{Tr}\left( \mathbf{V\Gamma }_{k,f}\mathbf{W}_k\mathbf{\Gamma }_{k,f}^{H} \right) \geqslant r_{k,f}^{\min}\left( I_{k,f}^{\mathrm{iner}}+I_{k,f}^{\mathrm{iter}}+\sigma ^2 \right),  
\end{equation}   
 where $I_{k,n}^{\mathrm{iner}}=a_{k,n}\mathrm{Tr}\left( \mathbf{V\Gamma }_{k,n}\mathbf{W}_k\mathbf{\Gamma }_{k,n}^{H} \right) 
  $, $I_{k,f}^{\mathrm{iner}}=a_{k,n}\mathrm{Tr}\left( \mathbf{V\Gamma }_{k,f}\mathbf{W}_k\mathbf{\Gamma }_{k,f}^{H} \right) 
  $ and $I_{k,i}^{\mathrm{iter}}=\sum_{\widetilde{k}\ne k}^K{\mathrm{Tr}\left( \mathbf{V\Gamma }_{k,i}\mathbf{W}_{\widetilde{k}}\mathbf{\Gamma }_{k,i}^{H} \right)}
  $, $r_{k,i}^{\min}=2^{R_{k,i}^{\min}}-1
  $, $i\in \left\{ n,f \right\}$ , $k\in \mathcal{K}$.
 
 % It is noted that the objective function of problem~\eqref{OP_MM} is non-smooth, 
 Furthermore, we introduce an auxiliary variable $\chi$ to transform the nonsmooth objective function in problem~\eqref{OP_MM} into a smooth one. Finally, the original problem~\eqref{OP_MM} can be recast equivalently as follows: 
\begin{subequations}\label{OP_MM_SDP_smooth}
	 	\setlength{\abovedisplayskip}{0pt}
	 	\setlength{\belowdisplayskip}{0pt}
	\begin{align}
		&\underset{\chi >0,\mathbf{a }_k,\mathbf{W}_k\succcurlyeq 0,\mathbf{V}\succcurlyeq 0}{\max}\chi , \\
		&s.t.~\text{Tr}\left[ \mathbf{V\Upsilon }_q\left( \sum_{k=1}^K{\mathbf{W}_k}  \right) \mathbf{\Upsilon }_{q}^{H} \right]  \geqslant \chi,\label{OP_MM_SDP_smooth:b} \\
		&   \ \ \ \ \  \sum_{k=1}^K{\text{Tr}\left( \mathbf{W}_k \right)}\leqslant P_{\max}
		, \label{OP_MM_SDP_smooth:c}  \\	
		&   \ \ \ \ \  \left[ \mathbf{V} \right] _{m,m}=1 , \label{OP_MM_SDP_smooth:e}\\
		&   \ \ \ \ \  \text{rank}\left( \mathbf{W}_k \right) =1,  \label{OP_MM_SDP_smooth:f} \\
		&   \ \ \ \ \  \text{rank}\left( \mathbf{V} \right) =1, \label{OP_MM_SDP_smooth:g}	 \\
		&   \ \ \ \ \  \eqref{OP_MM:d},~\eqref{R11 constrait},~\eqref{R21 constrait},~\eqref{R22 constrait} , 
	\end{align} 
\end{subequations}  
 where $i\in \left\{ n,f \right\}, k\in \mathcal{K}$, $m\in \mathcal{M}$, $q\in \mathcal{Q}$.
 
\subsection{Joint Active Beamforming And Power Allocation Coefficients Optimization}
In this subsection, we aim to optimize the active beamforming and power allocation coefficients by solving problem~\eqref{OP_MM_SDP_smooth} with given passive beamforming matrix $\mathbf{V}$. In particular, we can obtain the optimal $\left\{ \mathbf{W}_k \right\} $ and $\left\{ \mathbf{a}_k \right\} $ by solving the following optimization problem:
\begin{subequations}\label{AB_Max_SDP}
 	\setlength{\abovedisplayskip}{0pt}
 	\setlength{\belowdisplayskip}{0pt}
	\begin{align}
		&\underset{\chi >0,\mathbf{a }_k,\mathbf{W}_k\succcurlyeq 0}{\max}\chi , \\
		&s.t.~\eqref{OP_MM:d},~\eqref{R11 constrait},~\eqref{R21 constrait},~\eqref{R22 constrait},~\eqref{OP_MM_SDP_smooth:b},~\eqref{OP_MM_SDP_smooth:c},~\eqref{OP_MM_SDP_smooth:f}.  
	\end{align} 
\end{subequations}  
 
The difficulty to solve problem~\eqref{AB_Max_SDP} is the nonconvex constraints~\eqref{R11 constrait},~\eqref{R21 constrait},~\eqref{R22 constrait}, and~\eqref{OP_MM_SDP_smooth:f}. We first handle constraint~\eqref{R11 constrait} by introducing a new slack variable $\eta _k$, then we have
\begin{numcases}{}
		a_{k,n}\mathrm{Tr}\left( \mathbf{W}_k\mathbf{H}_{k,n} \right) \geqslant \eta _{k}^{2}, \label{slack_1}\\ 
		\eta _{k}^{2}\geqslant r_{k,n}^{\min}\left( I_{k,n}^{\mathrm{iter}}+\sigma ^2 \right) ,\label{slack_2}
\end{numcases}
where $\mathbf{H}_{k,n}=\mathbf{\Gamma }_{k,n}^{H}\mathbf{V\Gamma }_{k,n}
$ and $I_{k,n}^{\mathrm{iter}}=\sum_{\widetilde{k}\ne k}^K{\mathrm{Tr}\left( \mathbf{W}_{\widetilde{k}}\mathbf{H}_{k,n} \right)}$.% $i\in \left\{ n,f \right\}$.

Then, by applying the Schur complement theory~\cite{Xie2021JointOO}, \eqref{slack_1} can be
expressed in linear matrix inequality form
\begin{equation}\label{LMI}
\left[ \begin{matrix}
	a_{k,n}&		\eta _k\\
	\eta _k&		\mathrm{Tr}\left( \mathbf{W}_k\mathbf{H}_{k,n} \right)\\
\end{matrix} \right] \succcurlyeq 0,\exists \eta _k> 0
\end{equation} 

Furthermore, by using the successive convex approximation (SCA) approach based on First-order Taylor approximation, \eqref{slack_2} can be approximated as followed:
\begin{equation}\label{AB_Max_SDP:c21}
	\widetilde{\eta _{k}^{2}}+2\widetilde{\eta _k}\left( \eta _k-\widetilde{\eta _k} \right) \geqslant r_{k,n}^{\min}\left( I_{k,n}^{\mathrm{iter}}+\sigma ^2 \right),
\end{equation} 
where $\widetilde{\eta }_{k}$ is a fixed point and can be updated by $\widetilde{\eta }_{k}^{\left( t_1 \right)}=\eta _{k}^{\left( t_1 \right)}$,  $t_1$ is the iteration index. 

Next, we tackle the constraints~\eqref{R21 constrait} and~\eqref{R22 constrait}. Substituting $a _{k,f}=1-a _{k,n}$ into~\eqref{R21 constrait} and~\eqref{R22 constrait}, we have:
\begin{equation}\label{AB_Max_SDP:d1}
	\frac{r_{k,f}^{\min}\left( \frac{\mathrm{Tr}\left( \mathbf{W}_k\mathbf{H}_{k,n} \right)}{r_{k,f}^{\min}}-I_{k,n}^{\mathrm{iter}}-\sigma ^2 \right)}{r_{k,f}^{\min}+1}\geqslant a_{k,n}\mathrm{Tr}\left( \mathbf{W}_k\mathbf{H}_{k,n} \right) ,
\end{equation} 
\begin{equation}\label{AB_Max_SDP:e1}
	\frac{r_{k,f}^{\min}\left( \frac{\mathrm{Tr}\left( \mathbf{W}_k\mathbf{H}_{k,f} \right)}{r_{k,f}^{\min}}-I_{k,f}^{\mathrm{iter}}-\sigma ^2 \right)}{r_{k,f}^{\min}+1}\geqslant a_{k,n}\mathrm{Tr}\left( \mathbf{W}_k\mathbf{H}_{k,f} \right),
\end{equation} 
where $\mathbf{H}_{k,f}=\mathbf{\Gamma }_{k,f}^{H}\mathbf{V\Gamma }_{k,f}
$ and $I_{k,f}^{\mathrm{iter}}=\sum_{\widetilde{k}\ne k}^K{\mathrm{Tr}\left( \mathbf{W}_{\widetilde{k}}\mathbf{H}_{k,f} \right)}$.% $i\in \left\{ n,f \right\}$.

We note that the functions $a _{k,n}\text{Tr}\left( \mathbf{W}_k\mathbf{H}_{k,n} \right) 
$ and $a _{k,n}\text{Tr}\left( \mathbf{W}_k\mathbf{H}_{k,f} \right)$ in~\eqref{AB_Max_SDP:d1} and~\eqref{AB_Max_SDP:e1} are nonconvex. To deal with these non-convexity, we adopt the arithmetic-geometric mean  inequality~\cite{Sun2018JointBA}. Specifically, $a _{k,n}\text{Tr}\left( \mathbf{W}_k\mathbf{H}_{k,n} \right) 
$ and $a _{k,n}\text{Tr}\left( \mathbf{W}_k\mathbf{H}_{k,f} \right)$ can be approximated as 
\begin{equation}\label{CUB1}
	\setlength{\belowdisplayskip}{0pt}
	a_{k,n}\mathrm{Tr}\left( \mathbf{W}_k\mathbf{H}_{k,n} \right) \leqslant \frac{\beta _{k,1}a_{k,n}^{2}}{2}+\frac{\left( \mathrm{Tr}\left( \mathbf{W}_k\mathbf{H}_{k,n} \right) \right) ^2}{2\beta _{k,1}}\triangleq \mathcal{T} _{k,1},
\end{equation} 
\begin{equation}\label{CUB2}
	\setlength{\abovedisplayskip}{0pt}
	a_{k,n}\mathrm{Tr}\left( \mathbf{W}_k\mathbf{H}_{k,f} \right) \leqslant \frac{\beta _{k,2}a_{k,n}^{2}}{2}+\frac{\left( \mathrm{Tr}\left( \mathbf{W}_k\mathbf{H}_{k,f} \right) \right) ^2}{2\beta _{k,2}}\triangleq \mathcal{T} _{k,2},
\end{equation} 
where $\beta _{k,1}$ and $\beta _{k,2}$ are fixed points. The equality in~\eqref{CUB1} and~\eqref{CUB2} will always hold if $\beta _{k,1}=\frac{\text{Tr}\left( \mathbf{W}_k\mathbf{H}_{k,n} \right)}{a _{k,n}}$ and $\beta _{k,2}=\frac{\text{Tr}\left( \mathbf{W}_k\mathbf{H}_{k,f} \right)}{a _{k,n}}
$.

Based on the aforementioned transformations and approximations, constraints given in~\eqref{AB_Max_SDP:d1} and~\eqref{AB_Max_SDP:e1} can be approximated as follows
\begin{equation}\label{AB_Max_SDP:d11}
	 \setlength{\abovedisplayskip}{0pt}
	\setlength{\belowdisplayskip}{0pt}
	\frac{r_{k,f}^{\min}\left( \frac{\mathrm{Tr}\left( \mathbf{W}_k\mathbf{H}_{k,n} \right)}{r_{k,f}^{\min}}-I_{k,n}^{\mathrm{iter}}-\sigma ^2 \right)}{r_{k,f}^{\min}+1}\geqslant \mathcal{T} _{k,1},
\end{equation} 
\begin{equation}\label{AB_Max_SDP:e11}
 	\setlength{\abovedisplayskip}{0pt}
	\setlength{\belowdisplayskip}{0pt}
	\frac{r_{k,f}^{\min}\left( \frac{\mathrm{Tr}\left( \mathbf{W}_k\mathbf{H}_{k,f} \right)}{r_{k,f}^{\min}}-I_{k,f}^{\mathrm{iter}}-\sigma ^2 \right)}{r_{k,f}^{\min}+1}\geqslant \mathcal{T} _{k,2},
\end{equation} 

 %The final difficulty to solve problem~\eqref{AB_Max_SDP} is the rank-one constraint~\eqref{OP_MM_SDP_smooth:f}. 
 Finally, let us turn our attention to the rank-one constraint~\eqref{OP_MM_SDP_smooth:f}.  
 To address this issue, we exploit the semidefinite relaxation (SDR) technique by removing the rank-one constraints from the problem formulation. As a result, the relaxed problem is give as
 \begin{subequations}\label{AB_Max_SDR}
 	 	\setlength{\abovedisplayskip}{0pt}
  	\setlength{\belowdisplayskip}{0pt}
 	\begin{align}
 		&\underset{\chi >0,0<a_{k,n}<1,\mathbf{W}_k\succcurlyeq 0}{\max}\chi 
 		 , \\
 		&s.t.~\eqref{OP_MM_SDP_smooth:b},~\eqref{OP_MM_SDP_smooth:c},~\eqref{LMI},~\eqref{AB_Max_SDP:c21}, ~\eqref{AB_Max_SDP:d11},~\eqref{AB_Max_SDP:e11}. 
 	\end{align} 
 \end{subequations}   
  
  In the following theorem, we will verify the tightness of the SDR problem.
  \begin{theorem}\label{Rank1}
  	If the relaxed problem~\eqref{AB_Max_SDR} is feasible, then the solutions $\left\{ \mathbf{W}_k \right\} $ obtained by solving the problem~\eqref{AB_Max_SDR} always satisfy $\text{rank}\left( \mathbf{W}_k \right) \leqslant 1$, $k\in \mathcal{K}$.
  	
  	\textit{Proof}: Similar proof can be found in~\cite{Sun2018JointBA}. 
 \end{theorem}

Theorem~\ref{Rank1} represents the fact that we can obtain the rank-one solutions of problem~\eqref{AB_Max_SDP} by solving the convex problem~\eqref{AB_Max_SDR}.

To facilitate the understanding of the proposed algorithm, we
summarize it in \textbf{Algorithm~\ref{ABPAC}}.
\begin{algorithm}
	\caption{Proposed joint active beamforming and power allocation coefficients optimization algorithm}
	\label{ABPAC}
	\begin{algorithmic}[1]
		\STATE  \textbf{Initialize} $\beta _{k,1}^{\left( 0 \right)},\beta _{k,2}^{\left( 0 \right)},\widetilde{\eta }_{k}^{\left( 0 \right)},k\in \mathcal{K}
		$, and set ${t_1} = 0$.
		\REPEAT
		\STATE   ${t_1} = {t_1} + 1$;
		\STATE   update $ \mathbf{W}_{k}^{\left( t_1 \right)}$, $a_{k,n}^{\left( t_1 \right)}$ and $  \eta _{k}^{\left( t_1 \right)}$ by solving problem~\eqref{AB_Max_SDR} with given $\beta _{k,1}^{\left( t_1-1 \right)},\beta _{k,2}^{\left( t_1-1 \right)},\widetilde{\eta }_{k}^{\left( t_1-1 \right)}$;
		\STATE  update  $\widetilde{\eta }_{k}^{\left( t_1 \right)}=\eta _{k}^{\left( t_1 \right)}$, $\beta _{k,1}^{\left( t_1 \right)}=\frac{\text{Tr}\left( \mathbf{W}_{k}^{\left( t_1 \right)}\mathbf{H}_{k,n} \right)}{a _{k,n}^{\left( t_1 \right)}}$ and $\beta _{k,2}^{\left( t_1 \right)}=\frac{\text{Tr}\left( \mathbf{W}_{k}^{\left( t_1 \right)}\mathbf{H}_{k,f} \right)}{a _{k,n}^{\left( t_1 \right)}} $;
		\UNTIL {the objective value of problem~\eqref{AB_Max_SDR} converge}.
		\STATE  \textbf{Output}: $\mathbf{W}_{k} $ and $\mathbf{a}_{k}$, $ k\in \mathcal{K}$.
	\end{algorithmic}
\end{algorithm}
 \vspace{-0.2cm}
\subsection{Passive Beamforming Optimization} 
 For any given  $\left\{ \mathbf{W}_k \right\}$ and $\left\{ \textbf{a} _{k} \right\}$, the passive beamforming optimization problem is given by
 \begin{subequations}\label{PB_Max_SDP}
  	\setlength{\abovedisplayskip}{0pt}
  	\setlength{\belowdisplayskip}{0pt}
 	\begin{align}
 		&\underset{\chi >0,\mathbf{V}\succcurlyeq 0}{\max}\chi , \\
 		&s.t.~\eqref{R11 constrait},~\eqref{R21 constrait},~\eqref{R22 constrait},~\eqref{OP_MM_SDP_smooth:b},~\eqref{OP_MM_SDP_smooth:e},~\eqref{OP_MM_SDP_smooth:g} .	
 	\end{align} 
 \end{subequations} 
 
Now, the remaining non-convexity in problem~\eqref{PB_Max_SDP} lies in
the rank-one constraint~\eqref{OP_MM_SDP_smooth:g}. According to the sequential rank-one constraint relaxation (SRCR) algorithm~\cite{Cao2017ASC}, the constraint $\text{rank}\left( \mathbf{V} \right) =1
$ can be transformed equivalently as:
\begin{equation} \label{rank_ralex} 
	  	\setlength{\abovedisplayskip}{0pt}
	\setlength{\belowdisplayskip}{0pt} 
	\mathbf{e}_{\max}^{H}\left( \mathbf{V}^{\left( t_2 \right)} \right) \mathbf{Ve}_{\max}\left( \mathbf{V}^{\left( t_2 \right)} \right) \geqslant \varepsilon ^{\left( t_2 \right)}\mathrm{Tr}\left( \mathbf{V} \right), 
\end{equation}   
where $\mathbf{V}^{\left( t_2 \right)}$ is the obtained solution in the $t_2$-th iteration, $\mathbf{e}_{\max}\left( \mathbf{V}^{\left( t_2 \right)} \right) $ is the eigenvector corresponding to the maximum eigenvalue of $\mathbf{V}^{\left( t _2 \right)}$, $\varepsilon ^{\left( t_2 \right)}\in \left[ 0,1 \right]$ is a relaxation parameter in the $t_2$-th iteration. We can increase $\varepsilon ^{\left( t_2 \right)}$
from 0 to 1 sequentially via iterations to gradually
approach a rank-one solution. After each iteration, the relaxation parameter can be updated as
\begin{equation} \label{update relaxation parameter} 
    \setlength{\abovedisplayskip}{0pt}
	\setlength{\belowdisplayskip}{0pt} 
    \varepsilon ^{\left( t_2+1 \right)}\longleftarrow \min \left( 1,\frac{\lambda _{\max}\left( \mathbf{V}^{\left( t_2 + 1\right)} \right)}{\mathrm{Tr}\left( \mathbf{V}^{\left( t_2 +1 \right)} \right)}+\rho ^{\left( t_2+1 \right)} \right) ,
\end{equation}  
where $\lambda _{\max}\left( \mathbf{V}^{\left( t_2 \right)} \right) $ is the largest eigenvalue of $\mathbf{V}^{\left( t_2 \right)}$ and $\rho ^{\left( t_2 \right)}$ denotes the step size.

 As a result, in the $t_2$-th iteration, the optimization problem that needs to be solved is given as follows
\begin{subequations}\label{PB_Max_SROCR}
	 	\setlength{\abovedisplayskip}{0pt}
	 	\setlength{\belowdisplayskip}{0pt}
	\begin{align}
		&\underset{\chi >0,\mathbf{V}\succcurlyeq 0}{\max}\chi , \\
		&s.t.~\eqref{R11 constrait},~\eqref{R21 constrait},~\eqref{R22 constrait},~\eqref{OP_MM_SDP_smooth:b},~\eqref{OP_MM_SDP_smooth:e},~\eqref{rank_ralex} .  	
	\end{align} 
\end{subequations}  
 
Problem~\eqref{PB_Max_SROCR} is a semidefinite program (SDP) problem and can be solved by the CVX tool~\cite{cvx}. The procedure for optimizing passive beamforming is sketched in \textbf{Algorithm~\ref{PB_algorithm}}.
\begin{algorithm}  
	\caption{Proposed passive beamforming optimization algorithm}
	\label{PB_algorithm}
	\begin{algorithmic}[1]
		\STATE  Initialize $\mathbf{V}^{\left( 0 \right)}
		$ and $\rho ^{\left( 0 \right)}$. Set $\varepsilon ^{\left( t _2 \right)}=0$ and  ${t_2} = 0$.
		\REPEAT
		\STATE  Solve problem~\eqref{PB_Max_SROCR} with $\left\{ \varepsilon ^{\left( t_2 \right)},\mathbf{V}^{\left( t_2 \right)} \right\} 
		$ to obtain $\mathbf{V}^*$;
		\STATE \textbf{if} problem~\eqref{PB_Max_SROCR} is solvable 
		\STATE ~~~Update $\mathbf{V}^{\left( t_2+1 \right)}=\mathbf{V}^*$; 
		\STATE ~~~Update $\rho ^{\left( t_2+1 \right)}=\rho ^{\left( 0 \right)}
		$;
		\STATE \textbf{else}  
		\STATE ~~~Update $\mathbf{V}^{\left( t_2+1 \right)}=\mathbf{V}^{\left( t_2 \right)}
		$;  
		\STATE ~~~update $\rho ^{\left( t_2+1 \right)}=\frac{\rho ^{\left( t_2 \right)}}{2}
		$;
		\STATE \textbf{end}
		\STATE  Update $\varepsilon ^{\left( t _2 + 1 \right)}$ via~\eqref{update relaxation parameter};			
		\STATE  Update ${t_2} = {t_2} + 1$;	
		\UNTIL {$\frac{\mathrm{Tr}\left( \mathbf{V}^{\left( t_2 \right)} \right)}{\lambda _{\max}\left( \mathbf{V}^{\left( t_2 \right)} \right)}$ is below a predefined threshold and the objective value of problem~\eqref{PB_Max_SROCR} converges.}
		\STATE   \textbf{Output}: $\mathbf{V}$.
	\end{algorithmic}
\end{algorithm} 	
 \vspace{-0.3cm}
 \subsection{Proposed Algorithm, Complexity and Convergence}
 To facilitate the understanding of the proposed algorithm
 for solving problem~\eqref{OP_MM_SDP_smooth}, we summarize it in \textbf{Algorithm~\ref{proposed low algorithm}}.  The convergence of \textbf{Algorithm~\ref{proposed low algorithm}} is analyzed as follows.~\textbf{Algorithm~\ref{ABPAC}} and ~\textbf{Algorithm~\ref{PB_algorithm}} converge to a KKT stationary solution of problem~\eqref{AB_Max_SDP} and problem~\eqref{PB_Max_SDP}, respectively, which has been proved in ~\cite{Sun2018JointBA} and ~\cite{Cao2017ASC}.
In addition, the objective value of problem~\eqref{OP_MM_SDP_smooth} is non-decreasing after each iteration and the radar beampattern gain is upper bounded. Therefore, the proposed algorithm is guaranteed to converge. The complexities of \textbf{Algorithm~\ref{ABPAC}} and \textbf{Algorithm~\ref{PB_algorithm}} are $\mathcal{O} \left( T_{1}^{\max}\left( \max \left\{ N_{\mathrm{cont}},N_{\mathrm{T}} \right\} ^4\sqrt{N_{\mathrm{T}}}\log \frac{1}{\varpi _1} \right) \right) $ and $\mathcal{O} \left( T_{2}^{\max}\left( \max \left\{N_{\mathrm{cont}}+1,M \right\} ^4\sqrt{M}\log \frac{1}{\varpi _2} \right) \right) $, respectively, 
  where $\varpi _1$ and $\varpi _2$ are the solution accuracies, $T_{1}^{\max}$ and $T_{2}^{\max}$ are the number of iterations, $N_{\mathrm{cont}}=3K+Q+1$. The whole complexity of \textbf{Algorithm~\ref{proposed low algorithm}} is $\mathcal{O} \left( T_{3}^{\max}\left[ \begin{array}{c}
  	T_{1}^{\max}\left( \max \left\{ N_{\mathrm{cont}},N_{\mathrm{T}} \right\} ^4\sqrt{N_{\mathrm{T}}}\log \frac{1}{\varpi _1} \right) +\\
  	T_{2}^{\max}\left( \max \left\{N_{\mathrm{cont}},M \right\} ^4\sqrt{M}\log \frac{1}{\varpi _2} \right)\\
  \end{array} \right] \right)
  $, where $T_{3}^{\max}$ is the number of iterations for \textbf{Algorithm~\ref{proposed low algorithm}}.
 \begin{algorithm}
 	\caption{Proposed joint optimization algorithm}
 	\label{proposed low algorithm}
 	\begin{algorithmic}[1]
 		\STATE  Initialize $\mathbf{V}^{\left( 0 \right)}$ and set ${t_3} = 0$.
 		\REPEAT
 		\STATE  ${t_3} = {t_3} + 1$;
 		\STATE  update $\mathbf{W}_{k}^{\left( t_3 \right)} $ and $\mathbf{a}_{k}^{\left( t_3\right)}$ by \textbf{Algorithm~\ref{ABPAC}} with $\mathbf{V}^{\left( t_3-1 \right)}$; 
 		\STATE  update $\mathbf{V}^{\left( {t_3 } \right)}$ by \textbf{Algorithm~\ref{PB_algorithm}} with $\mathbf{W}_{k}^{\left( t_3 \right)} $ and $\mathbf{a}_{k}^{\left( t_3\right)}$;
 		\UNTIL {the objective value of problem~\eqref{OP_MM_SDP_smooth} converges.}
 		\STATE   \textbf{Output}: $\mathbf{W}_{k} $, $\mathbf{a}_{k}$, and $\mathbf{V}$, $ k\in \mathcal{K}$.
 	\end{algorithmic}
 \end{algorithm} 
   \vspace{-0.5cm}
  \section{Simulation Results} 
% In the experiments reported in this section, we use the following settings: 
\begin{figure}[t]
	 \setlength{\belowcaptionskip}{-15pt}
	\centering
	\includegraphics[scale=0.15]{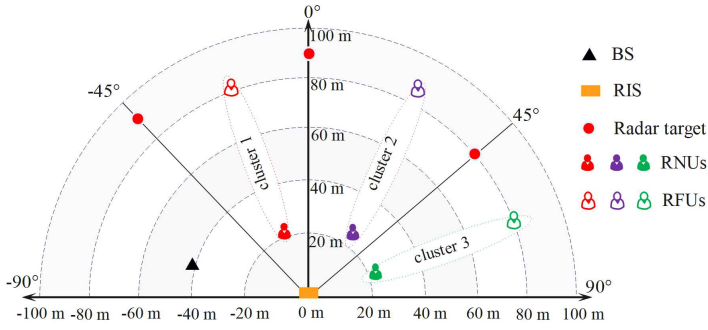}
	\caption{The simulated RIS-NOMA-ISAC system geometry}
	\label{simulation_fig}
\end{figure}

In this section, the performances of the proposed algorithm
for RIS-NOMA-ISAC system are evaluated through numerical simulations.
The simulated RIS-NOMA-ISAC system geometry is shown in Fig.~\ref{simulation_fig}. We assume that there are three clusters and three radar targets. The RIS is located in the origin of the coordinate, while the BS is located at $\left( -40, 10\right) $ meter $\left( \text{m} \right) $. In each cluster, the RNU and the RFU are randomly distributed on the half circles centered at $\left( 0,0 \right) \text{m}$ with radius of $r_{k,n}$ and $r_{k,f}$, respectively, where $r_{k,n}\in \left[ 20,25 \right]\mathrm{m} $ and $r_{k,f}\in \left[ 80,85 \right] \mathrm{m}$.  Let $\theta _{k,i}$ denote the angle from the RIS to user $\mathcal{U}\left( k,i \right)$ and we further assume that $\left\{ \theta _{1,i} \right\} $ in cluster 1, $\left\{ \theta _{2,i} \right\} $ in cluster 2, and $\left\{ \theta _{3,i} \right\} $ in cluster 3, are randomly distributed in the angle ranges $\left( -30^{\text{o}},-20^{\text{o}} \right] $, $\left(20^{\text{o}},30^{\text{o}} \right] 
$, and $\left( 60^{\text{o}},70^{\text{o}} \right] $, respectively, $i\in \left\{ n,f \right\} 
$. Without loss of generality, let $\theta _{k,n}=\theta _{k,f}$, $k\in \mathcal{K} 
$. The angle from the RIS to the three targets are set to $-45^{\text{o}}$,~$0^{\text{o}}$ and~$45^{\text{o}}$, and the corresponding radius are set to $90~\text{m}$, $90~\text{m}$ and $80~\text{m}$, respectively. With the angles of sensing targets, the desired beampattern can be defined as   
\begin{equation}\label{desired_beampattern}
	\mathcal{P}\left( \theta \right) =\left\{ \begin{array}{c}
		1, \theta _{\text{T}}-\frac{\Delta \theta}{2}\leqslant \theta \leqslant \theta _{\text{T}}+\frac{\Delta \theta}{2},\theta _{\text{T}}\in \left\{ -45^{\text{o}},0^{\text{o}},45^{\text{o}} \right\},\\
		0, \text{otherwise},~~~~~~~~~~~~~~~~~~~~~~~~~~~~~~~~~~~~~~~~~\\
	\end{array} \right. 
\end{equation}
where $\Delta \theta =6^{\text{o}}$ is the desired beam width, $\theta$ is the element of angle grid $\left[ -\frac{\pi}{2}:\frac{\pi}{100}:\frac{\pi}{2} \right] $, and the interested sensing angle set can be defined as $\mathcal{Q} _{\theta}=\left\{ \theta |\theta _{\mathrm{T}}-\frac{\Delta \theta}{2}\leqslant \theta \leqslant \theta _{\mathrm{T}}+\frac{\Delta \theta}{2},\theta _{\mathrm{T}}\in \left\{ -45^{\mathrm{o}},0^{\mathrm{o}},45^{\mathrm{o}} \right\} \right\}$.

All the channels follow the Rician fading, which can be modeled as in~\cite{Li2021JointBD}. The path loss at a reference distance of one meter is set to $-30{\text {~dB}}$, the path loss exponents are set to 2.2, the Rician factors are set to 3, the noise power is set to $-90\text{~dBm}$, the normalized spacing between two adjacent antennas(elements) is set as $\frac{d}{\lambda}=0.5$. The minimum QoS requirement for RNUs and RFUs are set to $R_{k,n}^{\min}=0.5$ bits/s/Hz and $R_{k,f}^{\min}=0.1$ bits/s/Hz, respectively. The maximum transmit power is set to $35$ dBm.
 \begin{figure}[t]
 	 \setlength{\abovecaptionskip}{3pt}
 	 \setlength{\belowcaptionskip}{-15pt}
 	\centering
 	\includegraphics[scale=0.4]{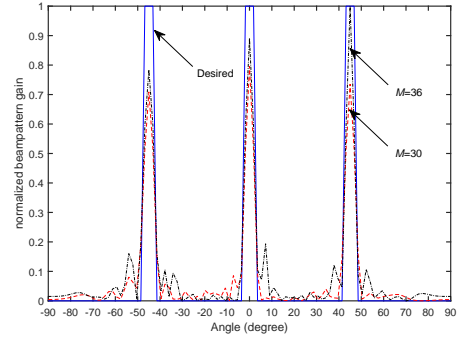}
 	\caption{Obtained beampattern gain with $N_{\text{T}} =9$}
 	\label{Beampattern_angle}
 \end{figure}
 \begin{figure}[t]
 	 \setlength{\abovecaptionskip}{3pt}
 	 \setlength{\belowcaptionskip}{-20pt}
 	\centering
 	\includegraphics[scale=0.4]{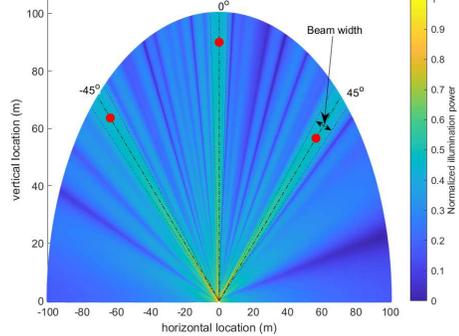}
 	\caption{Illumination power of the RIS with $N_{\text{T}} =9$ and $M=30$.}
 	\label{illumiate_power}
 \end{figure} 

In Fig.~\ref{Beampattern_angle}, we plot the normalized beampattern gain obtained over one random channel realization. The desired beampattern gain is obtained via~\eqref{desired_beampattern}. As illustrated in Fig.~\ref{Beampattern_angle}, the proposed scheme can achieve the dominant peaks of the beampattern gain in the angles of interest, i.e., $-45^{\text{o}}$,~$0^{\text{o}}$ and~$45^{\text{o}}$. Moreover, at the target directions, the achievable beampattern gains for $M=36$ are always higher than that for $M=30$, which implies that the performance of the proposed scheme can be improved by increasing the number of the RIS elements. In Fig.~\ref{illumiate_power}, we show that the normalized illumination power~\cite{Sankar2022BeamformingII} of the RIS on different angles and locations over one random channel realization. From Fig.~\ref{illumiate_power}, we can see that there are three brightest region towards the target directions. This is intuitive since the RIS attempts to steer both the active and passive beamforming towards the targets to maximize the minimum beampattern.

\begin{figure}[t]
	 \setlength{\belowcaptionskip}{-25pt}
	\centering
	\includegraphics[scale=0.4]{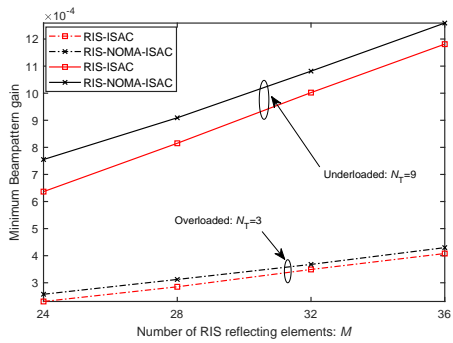}
	\caption{Obtained minimum beampattern gain by different schemes}
	\label{Fig5}
\end{figure}

\begin{figure}[t]
	 \setlength{\belowcaptionskip}{-25pt}
	\centering
	\includegraphics[scale=0.4]{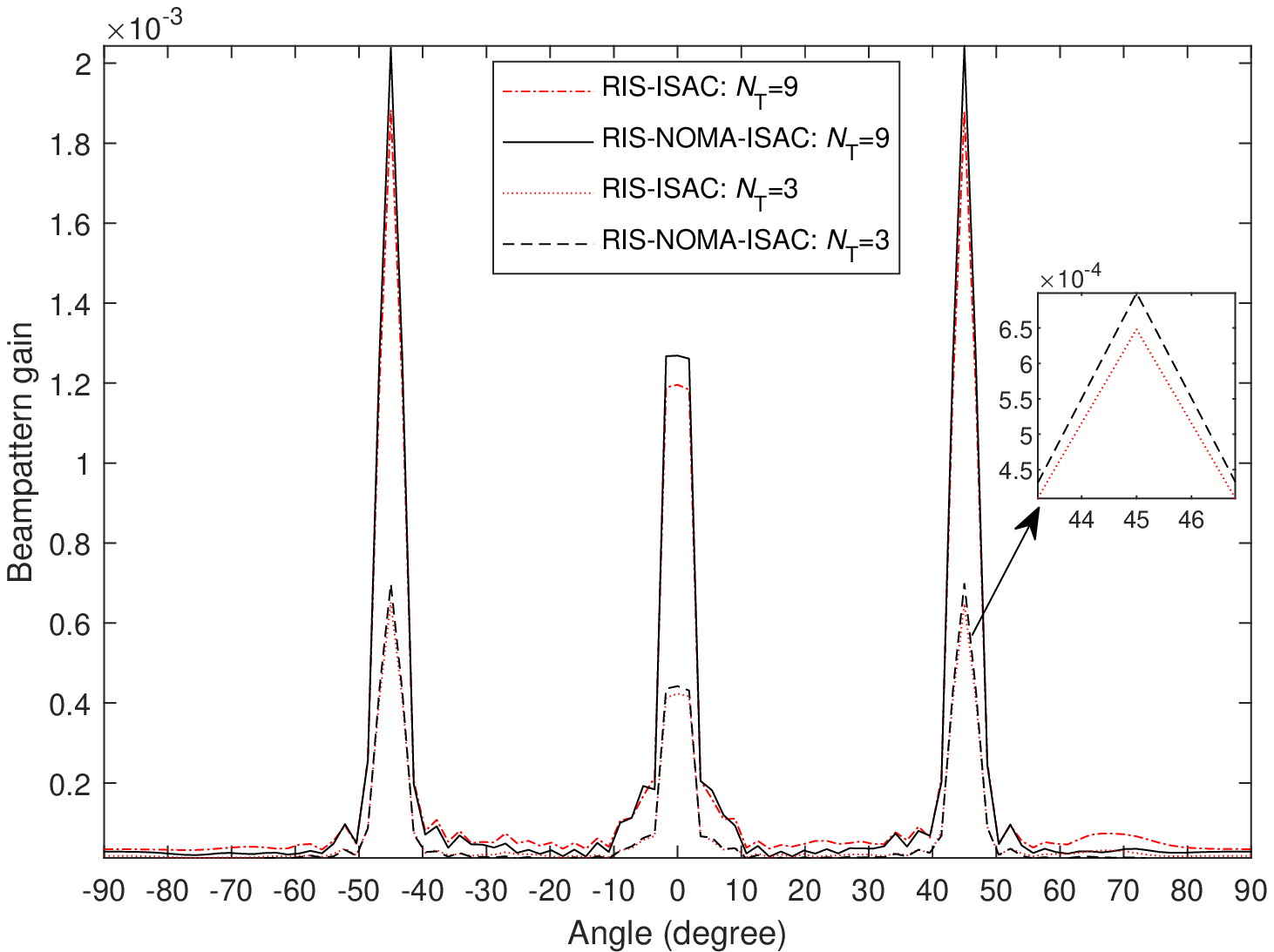}
	\caption{Obtained beampattern gain by different systems with $M=36$}
	\label{Fig6}
\end{figure}

To evaluate the performance of the proposed RIS-NOMA-ISAC system, the RIS assisted conventional ISAC (RIS-ISAC) system without NOMA is considered. The achievable rate of user $k$ can be obtained by: $R_k=\log _2\left( 1+\frac{\left| \mathbf{g}_{k}^{H}\mathbf{\Theta Gw}_k \right|^2}{\sum_{i\ne k}^{2K}{\left| \mathbf{g}_{i}^{H}\mathbf{\Theta Gw}_k \right|^2}+\sigma ^2} \right) 
 $, where $\mathbf{g}_k$ is the channel coefficients from the RIS to user $k$, $k\in \left\{ 1,2,\cdots ,2K \right\} $. The joint active and passive beamforming optimization problem for RIS-ISAC system can also be solved by our proposed algorithms. In Fig.~\ref{Fig5}, we compare the minimum beampattern gain obtained by the RIS-NOMA-ISAC and RIS-ISAC systems versus $M$. The results are obtained over 100 random channel realizations. First, it is observed that the achievable minimum beampattern gain monotonically increases with $M$. This is expected since more reflecting elements can provide higher possible beamforming gain towards the radar targets, thereby increasing beampattern gain. Second, the proposed RIS-NOMA-ISAC system outperforms the RIS-ISAC system under overloaded and underloaded cases. This is because the RIS-NOMA-ISAC system can mitigate the inter-user interference by SIC and provide more degrees of freedoms. for radar sensing. However, the RIS-ISAC system cannot well mitigated the inter-user interference. Fig.~\ref{Fig6} compares the beampattern gain obtained by different systems. We observe that the beampattern gain of all schemes have peaks towards target directions and our proposed RIS-NOMA-ISAC system has a stronger peak compared with the RIS-ISAC system. On the other hand, the proposed scheme can obtain higher beampattern gains at the worst angles. 
 \vspace{-0.2cm}
  \section{CONCLUSIONS}  
   \vspace{-0.2cm}
 We have proposed a novel RIS-NOMA-ISAC system, where the RIS is deployed to serve the communication users and assist radar sensing. A joint optimization problem over active beamforming, power allocation coefficients, and passive beamforming was formulated with the aim of maximizing the minimum beampattern gain. Due to the nonconvex nature of the formulated problem, an alternating optimization based algorithm was proposed to solve the original problem. Simulation results were presented to
 demonstrate the %effectiveness of the proposed algorithm and
  superiority of the proposed RIS-NOMA-ISAC system.

 \vspace{-0.2cm}
\bibliographystyle{IEEEtran}
 \bibliography{zjkbib}

\end{document}